# Advances in Run-Time Performance and Interoperability for the Adapteva Epiphany Coprocessor

David A. Richie[1] and James A. Ross[2]

[1]Brown Deer Technology, Forest Hill, MD
[2]U.S. Army Research Laboratory, Aberdeen Proving Ground, MD
drichie@browndeertechnology.com, james.a.ross176.civ@mail.mil

**Abstract**
The energy-efficient Adapteva Epiphany architecture exhibits massive many-core scalability in a physically compact 2D array of RISC cores with a fast network-on-chip (NoC). The architecture presents many features and constraints which contribute to software design challenges for the application developer. Addressing these challenges within the software stack that supports application development is critical to improving productivity and expanding the range of applications for the architecture. We report here on advances that have been made in the COPRTHR-2 software stack targeting the Epiphany architecture that address critical issues identified in previous work. Specifically, we describe improvements that bring greater control and precision to the design of compact compiled binary programs in the context of the limited per-core local memory of the architecture. We describe a new design for run-time support that has been implemented to dramatically improve the program load and execute performance and capabilities. Finally, we describe developments that advance host-coprocessor interoperability to expand the functionality available to the application developer.

## 1 Introduction

The Adapteva Epiphany MIMD architecture [1] is a scalable 2D array of RISC cores with minimal uncore functionality connected with a fast 2D mesh Network-on-Chip (NoC). Processors based on this architecture exhibit good energy efficiency and scalability via the 2D mesh network. The 16-core Epiphany III coprocessor has been integrated into the Parallella minicomputer platform where the RISC array is supported by a dual-core ARM CPU and asymmetric shared-memory access to off-chip global memory. Each of the 16 Epiphany III mesh nodes contains 32 KB of shared local memory (used for both program instructions and data), a mesh network interface, a dual-channel DMA engine, and a RISC

CPU core. Each RISC CPU core contains a 64-word register file, sequencer, interrupt handler, arithmetic logic unit, and a floating point unit. The 64-core Epiphany IV, fabricated on the 28 nm process, has demonstrated energy efficiency exceeding 50 GFLOPS per watt.

The Epiphany architecture remains a challenge to program. Contributing factors include the limited 32 KB local core memory, low off-chip bandwidth, and an unfamiliar and evolving proprietary software stack. This work focuses on development utilizing the vendor-provided Epiphany SDK (eSDK) and the COPRTHR SDK.

## 2 Background

Since the initial availability of the Epiphany-III coprocessor, software support for the Epiphany architecture has been available in the vendor-provided eSDK with GCC compiler support [2]. This SDK provides a basic interface to the coprocessor and includes library calls for the Epiphany cores and library calls for developing host code to interface to the Epiphany cores from the CPU host of the heterogeneous platform. The host API provides very typical interface calls for a coprocessor including the ability to open access to the coprocessor device, load and start a binary program, and read and write to memory accessible to the cores of the multi-core coprocessor. Although useful as a starting point for application development, the eSDK lacks basic support for a standard parallel programming model.

The COPRTHR SDK [3] was initially adapted to support the Epiphany architecture in order to provide support for OpenCL as a parallel programming API for Epiphany. This software stack was eventually refactored to provide a direct interface to Epiphany [4] providing more consistent semantics than those found in the eSDK as well as Pthreads support extended to a heterogeneous host-coprocessor platform. These features enabled the development of threaded MPI for Epiphany which provided the first demonstration of high performance benchmarks using a standard parallel programming API for Epiphany [5], [6]. Subsequently, this same software stack has supported the development of the ARL OpenSHMEM for Epiphany for which details will be reported elsewhere.

Threaded MPI, built upon the COPRTHR software stack, enabled a demonstration of multiple benchmarks including matrix-matrix multiply, 5-point stencil, 2D FFT, and N-body. In this prior work, nearly unmodified MPI code was compiled and executed on the Epiphany coprocessor. The most significant change required is the small addition of code at the beginning of the thread function to extract arguments passed in using Pthreads-style semantics. Additionally, a small host program is used to manage the coprocessor device and distributed memory transfers as well as to make the coprthr_mpiexec call to launch the parallel task. Despite this success notable issues and limitations were identified, and it is based on this experience and the analysis of the overall software stack that the present work has been motivated.

We report here on the design and implementation of improvements to the supporting software stack for Epiphany that both resolve some of the most serious issues, but also prepare for greater scalability for future 1,000+ core devices. Specific contributions of this work include 1) the reduction in size, and precise control over the binary program layout within the limited local core memory, 2) improvements in the functionality and performance of program loading and execution, and 3) greatly enhanced interoperability between the host platform and the Epiphany co-processor.

### 2.1 Analysis of the eSDK and COPRTHR-1 Software Stack

Each Epiphany core has 32 KB of local memory that must be shared for program instructions and local data, and the architecture has no hardware support for caching memory access to the larger shared global memory. This places a high premium on local core memory, differing substantially from the situation confronted with more conventional processor architectures. As such, the size of the binary program image becomes a critical factor in determining the available size of local data sets that can be processed, since the tradeoff is direct and unavoidable. As an example, for the threaded MPI implementation of the Cannon algorithm for matrix-matrix multiplication [5], 47% of this local memory per core was used for program instructions and control data. This limited the size of matrices held locally and also created a tradeoff in optimizations such as loop unrolling. An analysis of how this space was used revealed issues all the way down to the standard C runtime code generated by GCC. Whereas the generation of a few thousand bytes of unused instructions may go unnoticed for a conventional architecture where memory is measured in gigabytes, the inefficiency of the generated instructions becomes problematic under the size constraints of Epiphany local core memory.

Another issue identified in prior work was the performance of the program load and execute model based on calls provided in the eSDK. A program binary, compiled by GCC with Epiphany as a target, was either an SREC or ELF binary file. The eSDK loader performed a serial copy from the host to each core of the Epiphany coprocessor. Program execution could be automatic or deferred to a subsequent start call from the host. This design has many inefficiencies and created a situation where the execution of a kernel on the coprocessor required a complete load and restart each time. Experimental support for creating persistent threads existed within the COPRTHR-1 Pthreads layer that allowed creating a shared mutex, lockable by threads on either the host or coprocessor. However, setting up such a synchronization protocol is complex and only mitigates program load and restart when re-executing the same kernel from the host. The overall latency and overhead for the load and execution of an Epiphany binary was quite high with typical times on the order of tens of milliseconds for the ELF loader or hundreds of milliseconds for the SREC loader. This created a paradox for the platform insofar as the architecture was designed to support the accelerated execution of parallel kernels, whereas the run-time support for the loading and execution of these kernels was designed to load a full program image to the cores. A parallel fork and join model was impractical due to the substantial overhead of launching lightweight parallel kernels.

Interoperability between the host Linux platform and the Epiphany cores was quite limited. Whereas the Epiphany coprocessor provided fast parallel execution of small kernels, the ability to perform many supporting tasks like file IO, global memory allocation, and buffered print output was limited at best and typically inaccessible. By contrast, the host platform supporting execution of code on the coprocessor has access to the full features of a Linux operating system. Without interoperability support, the programmer developing code for the coprocessor found significant limitations compared to a typical development platform. It would be useful in some cases to have access to host calls directly, both system calls and user defined calls, and for many developers the lack of standard support for printf() hinders the ordinary code development and debugging process.

The eSDK host library calls used to interact with the coprocessor provided an opaque interface that also lacked good interoperability. Even with the improvements brought by the COPRTHR-1 direct API, operations as simple as copying data between host memory and coprocessor-accessible memory required specialized read/write functions with opaque memory objects. Although host and coprocessor have shared access to a configurable portion of the main global memory of the platform, each saw a different address space requiring translation. This resulted in unnecessary copying of data that could otherwise have been shareable if the software stack provided a unified virtual address space.

Finally, for some applications the requirement of employing a co-design methodology targeting the host and Epiphany cores separately becomes inconvenient. A simpler workflow would be the ability to write a program targeting the Epiphany coprocessor directly and produce a program binary, executable from the host, with all necessary host interaction abstracted away for the developer. In such cases it would be ideal to simply create a conventional C main program for the Epiphany processor and execute it directly. This type of capability, complementary to the co-design methodology, has been non-existent for this platform architecture.

A re-design of the COPRTHR-1 software stack was undertaken to address the issues identified above. Consistency has been maintained in that the previously developed threaded MPI layer runs more or less unmodified on this new COPRTHR-2 software stack, albeit with higher performance and increased support in functionality. From the programmer's perspective, the previous threaded MPI API remains unchanged, with the only modifications being ordinary compatibility issues within the MPI implementation layer.

# 3 Design

## 3.1 Challenges

Based on these lessons learned from the above described prior work, three overall objectives were identified for the re-design of the software stack. First, the size reduction and precise management and placement of executable instructions within the limited 32 KB of core-local memory needed to be re-examined and improved upon. Second, the performance and functionality of the program load and execution needed to be addressed, especially in view of scalability issues with the potential for this architecture to be scaled up to 1,000+ cores. Finally, the host-coprocessor interoperability required enhancements to make available greater functionality and a more integrated development environment for the overall platform. It is important to note that these objectives are interrelated, therefore, significant thought went into final solutions to address these objectives. As an example, increasing the interoperability could be achieved by developing library calls for execution on the coprocessor, but this would undermine the objective of minimizing the amount of memory used for executable instructions. This and many other tradeoffs necessitate lightweight solutions to each design challenge appropriate for the unique constraints of the Epiphany architecture.

## 3.2 Binary Program Layout

Examining the fully linked program binary produced by GCC targeting Epiphany revealed that GCC generated non-negligible code that was unused or ill-suited for the local core memory constraints of the architecture. This involved low-level code linked in as part of the generic GCC C run-time, in some cases to support features of C++ or JAVA, and more generally supporting the execution model of a program running on a conventional CPU. The code in question is known to be unused based on the COPRTHR compilation and run-time design. On a conventional platform where memory is measured in the gigabytes these routines would go without notice. However with a processor architecture that relies significantly on the careful management of 32 KB of local memory per core, the situation is quite different. Simply put, local memory is too precious to waste for unused or unnecessary routines.

In some cases, unused code was removed through modifications to the linker description file. Additionally, the lowest level of code supporting the C run-time was re-engineered starting from the point at which execution is started with a jump to the first startup routines executed prior to the main entry point of the compiled program. Further opportunities for minimizing program size were found by redesigning the main kernel launch code that was originally developed to support OpenCL, and was only adapted to support the Pthread-based execution model. This involved flattening the execution model to a 1D thread topology as opposed to supporting the n-dimensional range of OpenCL. Additionally, the OpenCL-style argument passing was replaced with the more efficient Pthread-style argument passing.

In the most basic terms, this initial effort involved stripping the code used to support core startup through the launch of application-specific kernels. The effort resulted in a reduction in the amount of memory used for instructions, but also provided a foundation for subsequent steps in the redesign of the C run-time model discussed below. For the example of the threaded MPI Cannon matrix-matrix multiple benchmark, the size of the program binary was reduced from 47% to 23% of the available 32 KB of local core memory.

## 3.3 Separation of system and application code

The program load and execute model supported by the eSDK, and utilized in the original COPRTHR-1 design, treated the compiled binary for a specific application kernel as an opaque program. This required a complete program load and restart each time a host program offloaded execution to the Epiphany cores. Although it was possible to create a persistent program that used synchronization protocols to interact with the host, this was left as a requirement for the application developer to create a custom solution. In an effort to improve the program load and execute performance, a design was chosen that separated core system code and application code to allow for the Epiphany cores to remain in a persistent state of execution. We call this common system code syscore. Although the facilities it provides, especially interoperability features described in greater detail below, begin to resemble a primitive operating system, we are reluctant to classify it as such at this time. The design of syscore is dominated by the need to be efficient yet occupy an extremely reduced memory footprint; these objectives are not always mutually consistent.

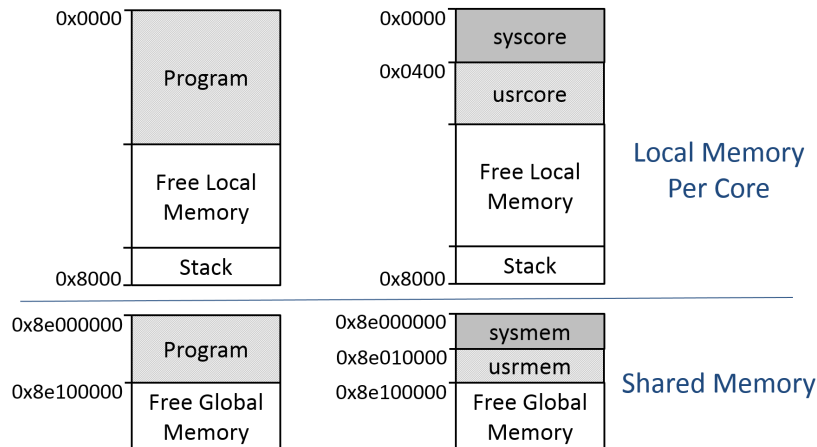

**Figure 1. Comparison of memory layout between the old monolithic program model and the new design that separates common system code rom applications specific user code.**

The idea of leaving the Epiphany cores in a persistent state of execution is to perform a "hot load" of application code which will be executed upon a signal from the host, and then upon completion the core is returned to a persistent wait state within *syscore*. In order to achieve this, the memory layout of the fully linked program binary is segmented to separate out user application code that may be safely copied into an actively executing core, provided that execution is safely held within the execution of system code.

A comparison of the memory layout is shown in Figure 1. The original layout contained executable sections in both the local core memory and global memory, in which common low-level system code and application code was comingled. The new layout uses a fixed segment for the core system code loaded into regions designated as syscore and sysmem within the local core memory and global shared memory, respectively. These segments are loaded once when the coprocessor is initialized. As long as execution for each core is held in a wait state within the syscore segment, the analogous segments containing application specific code, usrcore and usrmem, can be loaded safely. A signal from host to core will then cause execution to begin at the usrcore segment entry point to execute the application kernel.

The new design has several important features. The cost for this "hot load" will scale with the size of the application specific code only. Therefore it should be possible to achieve very low latency fork-and-join operations to support applications running on the host, thereby expanding the opportunities for taking advantage of the coprocessor to accelerate applications that fit this model. For scenarios where an application kernel must be re-executed there will be no need to reload the application code and all that will be required for re-execution is a signal from the host, further reducing overhead to utilize the coprocessor. This functionality transparently replaces the need for custom persistent thread solutions described above. The design also reduces the overhead for cycling through multiple application kernels by strictly scaling with the size of each kernel and not requiring a full restart of the core program.

An additional improvement was incorporated into the design that will be particularly beneficial if the Epiphany architecture is scaled to larger core counts. The original loader performed a serial load from the host which scaled with the number of cores, as shown in Figure 2. In the context of a 1,000+ core coprocessor, this scaling would create unacceptable overhead, and likely would be prohibitive in the practical use of such a coprocessor. For this reason a distributed tree-loader was incorporated into the syscore design which exhibits two important characteristics. First, it requires only one host-to-coprocessor copy of the loadable usrcore segment, specifically to core 0 in the scenario shown. (Loading the usrmem segment is unaffected by this design since it requires one copy operation from the host to shared global memory.) All subsequent copies of the loadable segment are made using on-chip inter-core copy operations, which is orders of magnitude higher bandwidth than off-chip copying. Second, the overall cost for the load will scale logarithmically, greatly reducing the overall load time.

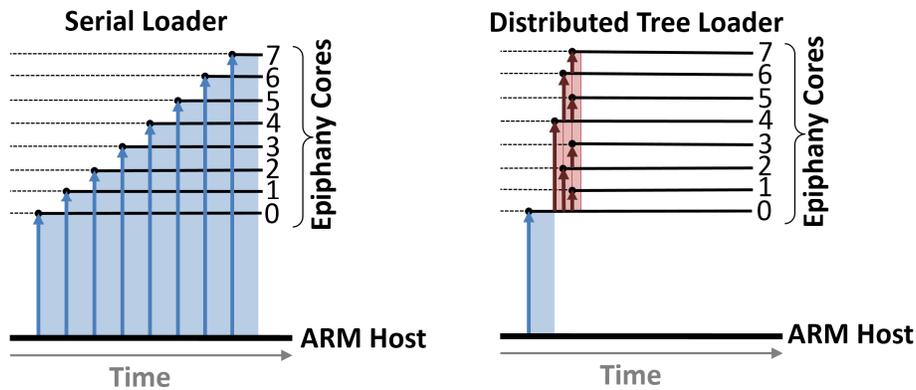

**Figure 2. Comparison of the original serial loader provided by the eSDK and the distributed tree loader implemented in the COPRTHR-2 run-time redesign.**

The impact on performance of the redesigned program load and execute model is significant. Table 1 provides timing measurements for various program load and execute scenarios. With the eSDK loader, the performance is limited by the low off-chip bandwidth (and file processing performance if using the SREC format). The program binary was serially copied for each core, which we've demonstrated to be inefficient within the context of a SPMD execution model.

**Table 1. Comparison of the cost for program load and execute times for the original eSDK loader, the new distributed tree loader, and additionally the cost to re-execute a previously loaded binary application program. All times use as an example the Canon matrix-matrix multiplication benchmark.**

| Procedure | Time (μs) |
|---|---|
| eSDK serial ELF loader | 73000 |
| COPRTHR-2 distributed tree loader | 2700 |
| COPRTHR-2 hot load and exec time for core 0 only | 790 |
| COPRTHR-2 re-execute | 40 |

## 3.4 Support for User-Defined Code Placement and Dynamic Calls

With the original eSDK, certain utility functions were deliberately placed in global memory rather than occupy space in local core memory. This was possible since the Epiphany coprocessor could execute instructions directly from global memory, albeit with a significant performance penalty. Suitable candidates for placement in global memory are functions executed infrequently and well outside critical loops. We follow this approach to minimize the footprint of executable instructions in local core memory with one important extension. Originally, the placement of functions in global memory was achieved by moving selected library code through specifications in the linker description file (LDF) used by the GCC linker for final linking and placement. Importantly, no mechanism existed for the application developer to selectively place code without modifying the LDF, a practice unfamiliar to most application developers.

We have introduced qualifiers to allow an application developer to selectively mark functions for placement in either local core memory or global memory, with a compile-time selectable default for all unmarked functions. This provides the developer with greater precision and control over the memory layout of an application binary. No special effort is required beyond marking a function with either __usrcore_call or __usrmem_call qualifiers to control placement. In theory, it should be possible to employ static analysis to fully or partially automate code placement, and this would be of interest to explore in future work. However, the use of a simple semantic enabling the programmer to mark-up source code to control the placement has two advantages. First, this will lead the programmer to structure an application in a way amenable for placement. As an example, it is advantageous to separate out one-time initialization code even if this would not be typically done for a conventional platform. Second, this approach exploits the programmers understanding of the application which can be sufficient, and often outperform automated techniques that may work in theory better than in practice. Providing a mechanism of expressibility now, in advance of research into more automated techniques, brings an immediate advantage as well.

More significantly, an alternative to the slow execution from global memory is introduced through dynamic calls. These are calls which reside in global memory but are copied into local core memory prior to execution. The use of dynamic calls was first attempted for Epiphany as a feature within GCC. However, we do not believe this feature was ever enabled by default, and we employ here a different approach to provide the developer with more control over the associated policies required to efficiently implement a dynamic call capability.

Functions marked with the __dynamic_call qualifier will be placed in global memory. In addition, an entry is created in a jump table that is automatically generated at compile time, along with an associated table storing the global address and size for each function marked as dynamic. The structure of the dynamic call (DC) jump table is shown in Figure 3. The design follows closely the technique used for dynamical loading of shared libraries. The first call to a dynamic function will be routed through the respective entry in the DC jump table. The structure of each entry is (initially) a branch to the next instruction, a mov of the call number into the inter-procedure scratch register ip, and then a branch to the DC loader. The DC loader uses the call number to lookup the address and size of the function stored in global memory so that it can be copied into a region set aside for dynamic calls. Once the instructions are copied, the first branch instruction for the associated entry in the DC jump table is directly modified to replace it with a branch to the location where the dynamic call has been loaded. Upon subsequent calls to this function, the only cost for being a dynamic call is a single branch indirection. The cost per dynamic call is 24 bytes of local core memory for each entry in the DC jump table.

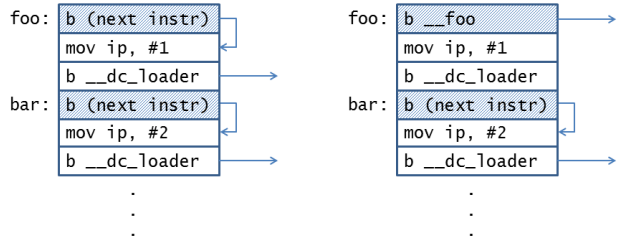

**Figure 3. Structure of the Dynamic Call (DC) jump table.** Initially dynamic calls are routed through the jump table and perform a branch to next instruction, load the inter-procedure scratch register with the call number, and then branch to the DC loader. The DC loader then uses the call number to look up the address and size of the function in global memory to copy the instructions into local core memory. Finally, the associated branch instruction in the DC jump table is replaced with a branch to the address where the function has been loaded into local core memory.

Developing a policy for managing the space set aside for dynamic calls, and specifically when certain calls should be invalidated, must be examined and developed in future work. At present, the developer can reset the DC jump table with a call that has the effect of invalidating all calls and freeing the set aside space. This limited capability still can be used effectively by the application developer for managing an application that might be broken up into stages that would otherwise consume too much local core memory if they were each loaded statically. Other candidates for dynamic calls are one-time initialization calls that typically perform significant bookkeeping operations. Upon startup, syscore itself uses such a dynamic call for initialization to allow for the size of the statically placed code to be minimized.

As a demonstration of the use of qualifiers for code placement and dynamic calls, selected layouts are generated for the Cannon matrix-matrix multiplication application code. Placing all code in core local memory provides the baseline reference and occupies 27% of the available core local memory. The memory use can be reduced to 18% by moving selected routines from the application and MPI library into global memory, with only a small change in execution time. However, if the inner-most matrix-matrix multiplication routine, representing the inner loop of the algorithm, is moved into global memory, the performance is significantly negatively impacted. By designating the routine as a dynamic call, most of the space-saving advantage is achieved (15% of core local memory is dedicated to application code). It should be noted that the dynamic call mechanism must copy the routine into local memory and therefore sufficient space is temporarily used. The advantage of the technique is that the same local memory can be used for other routines, although this aspect of the technique is not demonstrated in this example. Finally, the routines selected for placement are not the result of an exhaustive search for the ideal layout, but are intended to serve only as examples of placement.

**Table 2. Program size and execution time for selected layouts of the Cannon matrix-matrix multiply application using qualifiers to mark code for placement in global memory vs. core local memory and the use of dynamic calls**

| Layout | User Code | Time |
|---|---|---|
| Local memory only | 8736 bytes | 9.4 msec |
| Selected use of global memory[1] | 5960 bytes | 10.4 msec |
| Selected use of global memory[2] | 4864 bytes | 145.7 msec |
| Selected use of global memory with dynamic call[3] | 5064 bytes | 10.7 msec |

[1]MPI_Init(), MPI_Cart_create(), MPI_Cart_shift(), MatrixMultiplyComm() in global memory
[2]Same as [1] with MatrixMultiply() in global memory
[3]Same as [2] with MatrixMultiply() as dynamic call

## 3.5 Host-Coprocessor Interoperability

The Epiphany coprocessor is supported asymmetrically by a host processor forming a heterogeneous platform. In the case of Parallella this consists of an ARM host running Linux and tightly coupled with an Epiphany processor through a shared memory mapping. Although not exclusive, most applications using Epiphany will rely upon host code, therefore the potential for interoperability between the two architectures becomes an interesting question for investigation and an opportunity for development. Existing support for interoperability has been quite limited. Interactions with the coprocessor have employed an opaque interface typical of many offload accelerators. Conversely, the ability for code executing on the Epiphany cores to interact with the host system has been equally limited. As an example, support for printf() used by many developers as a critical debugging tool to provide insight into where and why a program may fail has been unsupported. Other utilities have also been missing such as file IO, for example. A more complex but critical example is the ability to allocate shared global memory from an Epiphany core, which could be done easily if a core had access to the COPRTHR host call, dmalloc(), that provides a full implementation of malloc for the shared global memory region. Interoperability issues were addressed in the redesign of the COPRTHR-2 software stacks, and are described here.

A significant development introduced for interoperability was a remapping on the Linux host of the memory address space shared with the Epiphany coprocessor. The eSDK provides opaque memory objects required for all memory read/write operations, which hinders interoperability. With a remapping of the address space we create a unified virtual address-space (UVA) that allows consistency between pointers used on the host and coprocessor. This development has the effect of replacing specialized memory copy calls provided by the eSDK with ordinary memcpy() calls on the host. The impact of a UVA becomes more pronounced with additional capabilities discussed below, for example it is because of the UVA that arguments to host system calls passed directly from Epiphany cores may be used without issue. Creating a UVA between host and coprocessor enables powerful capabilities for interoperability and tight integration of applications.

The coprocessor cores are best suited for performing critical loops with high floating point intensity, exemplifying the typical benefits of an offload accelerator. However, combined with the limited local core memory available for executable instructions, this has the effect of constraining the application space suitable for Epiphany. Even with dynamic calls extending the effective size of the code that can be efficiently executed, there are many supporting utilities that would either be efficient or simply convenient to execute outside of a critical loop. File IO is one an example; an efficient approach to performing operations as simple as opening, writing, and closing a file on the host Linux platform have remained inaccessible to the Epiphany cores. We address this limitation with the introduction of a *host call infrastructure* integrated into the compilation model and run-time support.

The idea of a host call is relatively simply in principle, and is similar to the concept of a remote procedure call (RPC). We would like to be able to execute a call on an Epiphany core that is carried out by proxy on the host, with all effects accurately reproduced. Design constraints are the desire to keep the executable program in local core memory small and avoid the implementation of bloated interface code. We present a design for host calls that cost only 8 bytes per call with a fixed cost of 128 bytes to provide the feature as a service within syscore. The design allows host system calls as well as user–defined functions to be executed indirectly from Epiphany cores.

Key to the design is an automatically generated host call jump table similar to the one used for dynamic calls. For each call, an entry is generated consisting of a mov instruction placing the call number into the inter-procedure scratch register, ip, followed by a branch to the code that implements the host callback mechanism (with entry labeled _ehostcall). For any host call, the call number, arguments passed by register, and current stack pointer are stored at a location accessible on the host. Then the low-level run-state value is modified by setting a high bit that the host daemon monitoring the execution of code on the coprocessor will identify as a request for host call execution. The core will then spin until signaled that the host call has been executed. Using the stored information, the host call is performed by routing through a call vector that connects the call number to the selected host function. By convention call numbers below 512 are interpreted as Linux system calls and dispatched directly. Call numbers between 512 and 1023 are reserved for utilities provided by the run-time. Finally, call numbers of 1024 or higher are interpreted as user-defined. Upon completion of the host call, the host daemon signals the core to proceed. It should be noted that here the creation of a UVA is critical in greatly simplifying argument passing, and allowing for data structures containing a "pointer to a pointer" to work without issue.

This infrastructure is powerful for extending the functionality available to the Epiphany cores by allowing them to interact nearly directly with the host Linux platform. It elegantly exposes Linux systems calls which can be used for file IO, so writing to files is now trivial. A user-defined function in a host application can also be exposed simply by marking it with a simple macro that registers the function at compile time, and all of the required linking is handled automatically making the call available from an Epiphany core. The overhead with the current implementation was measured using the wait time on the core to execute a user-defined host call that performed no operations, and found to be 41 μsecs.

## 4 Related Work

This work focuses on the software development support provided by the Adapteva Epiphany SDK (eSDK) and the COPRTHR SDK. Other methods investigated for programming Epiphany include Erlang [8], Array Manipulation Language [9], and CAL Actor Language with Network Language [10]. The Parallel Processing Group at the University of Ioannina has published new support OpenMP 4.0 offload support for the Parallella board using their OpenMP C Compiler [11]. This standard OpenMP model also removes the requirement for explicit host code, enabling improvements in productivity and software portability. With the exception of threaded MPI and ARL OpenSHMEM for Epiphany, additional programming models based on message passing or PGAS paradigms are not known to have been implemented.

## 5 Conclusions and Future Work

We have advanced the run-time performance and interoperability of the software stack supporting the Epiphany coprocessor platform. The redesign and implementation of improvements both resolve some of the most serious issues, but also prepare for greater scalability for future 1,000+ core devices. Specific contributions of this work include 1) the reduction in size, and precise control over the binary program layout within the limited local core memory, 2) improvements in the functionality and performance of program loading and execution, and 3) greatly enhanced interoperability between the host platform and the Epiphany co-processor.

In future work we will examine dynamic call policies to study their overall effect on performance and management of space used for loading dynamic calls into local core memory. In addition, characteristics of host call interoperability will be studied to better understand the impact on performance, especially in the context of a coprocessor scaled to 1,000+ cores.

## Acknowledgment

The authors wish to acknowledge the U.S. Army Research Laboratory-hosted Department of Defense Supercomputing Resource Center for its support of this work.